\documentclass[journal]{IEEEtran}
\usepackage{amsfonts}
\usepackage{bm}
\usepackage{amsmath,amsfonts,amsthm,amssymb}
\usepackage{setspace}
\usepackage{Tabbing}
\usepackage{fancyhdr}
\usepackage{lastpage}
\usepackage{extramarks}
\usepackage{chngpage}
\usepackage{algorithmic}
\usepackage{soul,color}
\usepackage{epsfig}
\usepackage{graphicx,float,wrapfig,subfigure}
\usepackage{dsfont}
\usepackage{longtable}
\usepackage{wrapfig}
\usepackage{stfloats}
\usepackage{cite}
\usepackage{graphicx}
\usepackage{array}
\usepackage{multirow}{\tiny }
\usepackage{multicol}
\usepackage{colortbl}
\usepackage{tabularx}
\usepackage{mdwmath}
\usepackage{mdwtab}
\usepackage{color}
\usepackage{verbatim}
\usepackage{amsmath}
\usepackage{tikz}
\usetikzlibrary{calc}
\usepackage{flowchart}
\usetikzlibrary{shapes.geometric, arrows}
\usepackage{overpic}
\usepackage{epstopdf}
\usepackage{url}
\usepackage[colorlinks,linkcolor=black,anchorcolor=black,citecolor=black,urlcolor=black]{hyperref} 

\usepackage{makecell}
\usepackage{textcomp}
\usepackage[ruled,linesnumbered,vlined]{algorithm2e}

\newfloat{model}{thp}{lop}

\usepackage{threeparttable}

\usepackage{color}
\makeatletter %
\let\myorg@bibitem\bibitem
\def\bibitem#1#2\par{%
	\@ifundefined{bibitem@#1}{%
		\myorg@bibitem{#1}#2\par
	}{%
		\begingroup
		\color{\csname bibitem@#1\endcsname}%
		\myorg@bibitem{#1}#2\par
		\endgroup
	}%
}

\makeatother %

\allowdisplaybreaks

\usepackage{footnote}
\makesavenoteenv{tabular}
\makesavenoteenv{table}

\begin{document}

	\title{
On the Choice of Loss Function \\
in Learning-based Optimal Power Flow
	}

\author{
\IEEEauthorblockN{Ge~Chen and Junjie~Qin}\\
\IEEEauthorblockA{\textit{Electrical and Computer Engineering, Purdue University}\\ West Lafayette, IN, USA \\
\texttt{\{chen4911, jq\}@purdue.edu}
}
}

	

\maketitle

\begin{abstract}
We analyze and contrast two ways to train machine learning models for solving AC optimal power flow (OPF) problems, distinguished with the loss functions used. The first trains a mapping from the loads to the optimal dispatch decisions, utilizing mean square error (MSE) between predicted and optimal dispatch decisions as the loss function. The other intends to learn the same mapping, but directly uses the OPF cost of the predicted decisions, referred to as \emph{decision loss}, as the loss function. 
In addition to better aligning with the OPF cost which results in reduced suboptimality, the use of decision loss can circumvent feasibility issues that arise with MSE when the underlying mapping from loads to optimal dispatch is discontinuous. 
{Since decision loss does not capture the OPF constraints, we further develop a neural network with a specific structure and introduce a modified training algorithm incorporating Lagrangian duality to improve feasibility.} This result in an improved performance measured by feasibility and suboptimality as demonstrated with an IEEE 39-bus case study. 

\end{abstract}

\begin{IEEEkeywords}
Optimal power flow, decision loss, machine learning, mean square error, Lagrangian duality.
\end{IEEEkeywords}

\section{Introduction} \label{sec_intro}

AC optimal power flow (OPF) is an essential tool for managing power system operations \cite{abdi2017review}. It allows operators to determine the most economical dispatch strategy while satisfying consumer demands and security constraints. With the rising integration of renewable resources, net loads may fluctuate rapidly \cite{olauson2016net}. Hence, there is a critical need to develop an approach that can solve AC OPF in real-time to follow these variations in net loads.

Traditionally, AC OPF is solved by the interior-point method. This method can provide local optima with guaranteed feasibility \cite{skolfield2022operations}. However, it may be computationally demanding, particularly for large-scale systems \cite{9612004}. To reduce the computational burden, researchers have proposed several linearization techniques, like DC OPF \cite{9205647} to simplify the problem. However, such techniques may affect both the optimality and feasibility of the solutions.
Advanced relaxation methods, such as second-order cone and semi-definite programming relaxations \cite{8636236}, have been proposed. While these relaxations can offer improved solutions, their computational complexities may still be high \cite{10058008}. As a result, there is a pressing need to develop novel approaches that can solve AC OPF in real time.

The development of smart meter technology has reduced the cost of collecting operational data from power systems. With this data, machine learning models can be trained to act as the surrogates of AC OPF \cite{9091534}. After training, these surrogates can predict optimal dispatching decisions in real time. For example, a multi-layer perceptron was trained to predict dispatch decisions in \cite{9844847}. {This multi-layer perceptron was replaced by a graph neural network in \cite{10023997} to enhance prediction accuracy. Considering that predicted decisions might not always satisfy power flow constraints, the Lagrangian duality was incorporated in the training process to improve feasibility \cite{9335481}.}

Most studies, such as \cite{9844847,10023997,9335481}, employ the mean square error (MSE) between actual and predicted optimal decisions as a loss function for training. The effectiveness of the MSE loss has been confirmed across various case studies. However, Elmachtoub et al. \cite{elmachtoub2022smart} pointed out that MSE may not accurately reflect decision quality. Although a perfect surrogate with zero MSE can yield optimal decisions, learning such a model is almost impossible in practice. Moreover, these works usually use local minima from the interior point method (IPM) as training labels. Due to the non-convex nature of OPF, small changes in loads can lead to different local minima becoming the solutions of IPM. Hence, the target mapping to be learned, i.e., the mapping from loads to these labels, may be discontinuous. However, these work often choose continuous neural network as their surrogates to learn this discontinuous mapping. This mismatch may compromise the feasibility of the trained surrogates.

{
Recently, ``decision loss" has emerged as a novel loss function for training machine learning models in decision-making tasks \cite{Wilder_Dilkina_Tambe_2019,NEURIPS2022_0904c7ed}. This loss utilizes the objectives of these tasks as the training loss functions and has demonstrated superior optimality compared to MSE. However, to the best of our knowledge, the application and effectiveness of decision loss in AC OPF problems have not yet been explored.}

{This paper introduces decision loss for training AC OPF surrogates and compares its effectiveness with the widely-used MSE. Our work makes three distinct contributions compared to existing research:}
\begin{enumerate}
\item We provide a thorough analysis of the optimality and feasibility issues introduced by using MSE. Several examples are also provided to illustrate these issues.
\item {We introduce a novel loss function, the decision loss, for training AC OPF surrogates. A detailed discussion is further presented to explain its advantages in addressing the issues caused by using MSE.}
\item We develop a specialized neural network and incorporate a Lagrangian duality-based training algorithm to improve the feasibility of predicted decisions.
\end{enumerate}
The remaining parts are organized as follows. Section \ref{sec_formulation} formulates the AC OPF model. Section \ref{sec_solution} introduces the drawbacks of the MSE loss and the formulation of the decision loss. Section \ref{sec_case} demonstrates simulation results, and Section \ref{sec_conclusion} concludes this paper.

\section{Problem Formulation} \label{sec_formulation}
AC OPF aims to find the best dispatch decision that can minimize the total generation cost of the whole power grid while satisfying users' demands and security constraints. By using $i \in \mathcal{N}$ and $(i,j) \in \mathcal{E}$ to represent the indexes of buses and lines, AC OPF can be formulated as
\begin{align}
\hspace{-2mm}\min_{\mathbf{p}^{\mathrm{g}}, \mathbf{q}^{\mathrm{g}}} \quad &\mathrm{Cost}( \mathbf{p}^{\mathrm{g}})  \tag{AC OPF} \label{eq:objective} \\
\text{s.t. } \quad &V_i^{\mathrm{min}} \leq V_i \leq V_i^{\mathrm{max}},  \ \ \quad  \qquad \qquad \quad \forall i \in \mathcal{N}, \label{eq:voltage_magnitude} \\
& p_i^{\mathrm{g,min}} \leq p^{\mathrm{g}}_i \leq p_i^{\mathrm{g,max}},  \ \ \  \qquad \qquad \quad \forall i \in \mathcal{N}, \label{eq:pg_limits} \\
& q_i^{\mathrm{g,min}} \leq q^{\mathrm{g}}_i \leq q_i^{\mathrm{g,max}}, \ \ \  \qquad \qquad \quad\forall i \in \mathcal{N}, \label{eq:qg_limits} \\
& |p^{\mathrm{f}}_{ij}|^2 + |q^{\mathrm{f}}_{ij}|^2 \leq s_{ij}^{\mathrm{max}}, \qquad \qquad \qquad \forall (i,j) \in \mathcal{E}, \label{eq:line_flow} \\
& p^{\mathrm{f}}_{ij} = g_{ij}V_i^2 - V_i V_j (g_{ij}\cos(\theta_i-\theta_j) \notag \\
& \quad \quad + b_{ij}\sin(\theta_i-\theta_j)), \qquad \qquad \quad   \forall (i,j) \in \mathcal{E}, \label{eq:pf_relation} \\
& q^{\mathrm{f}}_{ij} = -b_{ij}V_i^2 - V_i V_j (g_{ij}\sin(\theta_i-\theta_j) \notag \\
& \quad \quad - b_{ij}\cos(\theta_i-\theta_j)),  \qquad \qquad \quad \forall (i,j) \in \mathcal{E}, \label{eq:qf_relation} \\
& p^{\mathrm{g}}_i - p^{\mathrm{d}}_i = \sum_{j \in N} p^{\mathrm{f}}_{ij}, \ \ \ \  \qquad \qquad \qquad \forall i \in \mathcal{N}, \label{eq:active_balance} \\
& q^{\mathrm{g}}_i - q^{\mathrm{d}}_i = \sum_{j \in N} q^{\mathrm{f}}_{ij}, \ \ \ \  \qquad \qquad \qquad \forall i \in \mathcal{N}, \label{eq:reactive_balance}
\end{align}
where the objective $\mathrm{Cost}(\mathbf{p}^{\mathrm{g}})$ is to minimize the generation cost. Constraint \eqref{eq:voltage_magnitude} restricts the voltage magnitude at each bus $i$ between the lower bound $V_i^{\mathrm{min}}$ and upper bound $V_i^{\mathrm{max}}$, respectively. Constraint \eqref{eq:pg_limits} ensure the active generation at each bus within the  allowable range, where $p_i^{\mathrm{g,min}}$ and $p_i^{\mathrm{g,max}}$ represent the maximum and minimum allowable active generation. Similarly, constraint \eqref{eq:qg_limits} restricts the reactive generation with its operational bounds $q_i^{\mathrm{g,min}}$ to $q_i^{\mathrm{g,max}}$. The apparent power flow on each transmission line $(i,j)$, which includes both active $p^{\mathrm{f}}_{ij}$ and reactive $q^{\mathrm{f}}_{ij}$ components, should not exceed the upper limit $s_{ij}^{\mathrm{max}}$, which is described by constraint \eqref{eq:line_flow}. Constraints \eqref{eq:pf_relation} and \eqref{eq:qf_relation} are power flow equations, which define the relationship between the power flows, voltage magnitude $V_i$, and angle $\theta_i$ as well as the line conductance $g_{ij}$ and susceptance $b_{ij}$. Constraints \eqref{eq:active_balance} and \eqref{eq:reactive_balance} are the active and reactive power balance requirements on each bus. 

As mentioned in Section \ref{sec_intro}, AC OPF is usually tackled using IPM which may be time-consuming. Although some relaxation techniques have been proposed to simply the problem, they may still be computationally expensive.

\section{Solution Methodology} \label{sec_solution}
To improve computational efficiency, machine learning models can be trained to act as the surrogate of AC OPF. In this section, we first give an overview of the learning-based OPF methods. Then, we discuss the challenges caused by the MSE loss in detail. 
{We then outline the common procedure of training the surrogate using the MSE loss. Albeit simple, this approach may lead to optimality and feasibility issues.
After that, we introduce the decision loss. Although the decision loss can mitigate the aforementioned issues, it does not directly capture the OPF constraints and may result in infeasible solutions. Thus, we develop a specialized neural network and a Lagrangian duality-incorporated training to improve its feasibility.}

\subsection{Overview of learning-based OPF methods}

The learning-based OPF methods aims to train a surrogate of AC OPF that can directly predict the optimal dispatch decision for a given load condition. The input feature $\mathbf x$ is usually defined as the collection of active and reactive loads, while the output label $\mathbf y^*$ is the optimal dispatch decision:
\begin{align}
\mathbf x = (\mathbf p^\mathrm{d}, \mathbf q^\mathrm{d}),\quad \mathbf y^* = (\mathbf p^{\mathrm{g},*}, \mathbf q^{\mathrm{g},*}), \label{eqn_x}
\end{align}
where $\mathbf p^\mathrm{d}$ and $\mathbf q^\mathrm{d}$ are the vector forms of $p^\mathrm{d}_i$ and $q^\mathrm{d}_i$; $\mathbf p^\mathrm{g}$ and $\mathbf q^\mathrm{g}$ are the vector forms of $p^\mathrm{g}_i$ and $q^\mathrm{g}_i$. Here, we use superscript $*$ to denote the optimal solution. Then, the surrogate model $\bm \pi(\cdot)$ can be expressed as:
\begin{align}
\hat{\mathbf y} = \bm \pi(\mathbf x; \mathbf W), \label{eqn_model}
\end{align}
where $\hat{\mathbf y}$ represents the predicted decision; $\mathbf W$ is the parameters to be learned from historical data.

\subsection{MSE loss and associated challenges} \label{sec_MSE}
We can collect historical load conditions and corresponding optimal dispatch decisions, i.e., $\mathcal{D}=\{(\mathbf x_n, \mathbf y_n^*), \forall n \in D\}$, as training samples to train the surrogate $\bm \pi$. Most prior studies use the MSE between predicted and actual optimal decisions as the loss function for training:
\begin{align}
L^\mathrm{MSE}_\mathbf W(\mathbf y^*, \mathbf x) = \frac{1}{|D|}\sum_{n \in \mathcal{D}} \Vert \mathbf y^*_n - \bm \pi(\mathbf x_n; \mathbf W) \Vert^2, \label{eqn_MSE}
\end{align}
Although the effectiveness of MSE has been verified in various case studies, it may cause the following two issues:
\subsubsection{Optimality issue}
\begin{figure}
		\vspace{-4mm}
	\centering
	{\includegraphics[width=0.9\columnwidth]{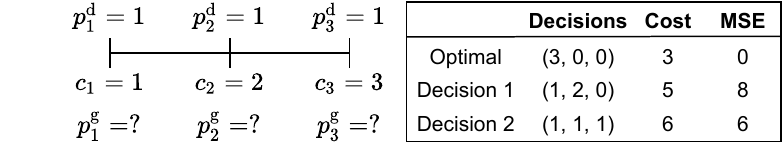}}
	\vspace{-4mm}
 	\caption{{An example where a decision with a high MSE exhibits a lower cost than another decision with a lower MSE.}}
	\label{fig_MSE_1}
	\vspace{-6mm}
\end{figure}

Elmachtoub et al. \cite{elmachtoub2022smart} has pointed out that MSE may not accurately measure the optimality of decisions. Fig. \ref{fig_MSE_1} presents an example to demonstrate this issue. In this example, a lossless 3-bus system is dispatched to maintain power balance with active loads of $(1,1,1)$ and unit generation costs of $(1,2,3)$. The optimal dispatch decision is $(3,0,0)$. However, the first candidate decision $(1,2,0)$, despite having a higher MSE, incurs a lower cost compared to the second decision $(1,1,1)$.

\subsubsection{Feasibility issue}
\begin{figure}
		\vspace{-4mm}
	\subfigbottomskip=-4pt
	\subfigcapskip=-4pt
	\centering
	\subfigure[]{\includegraphics[width=0.8\columnwidth]{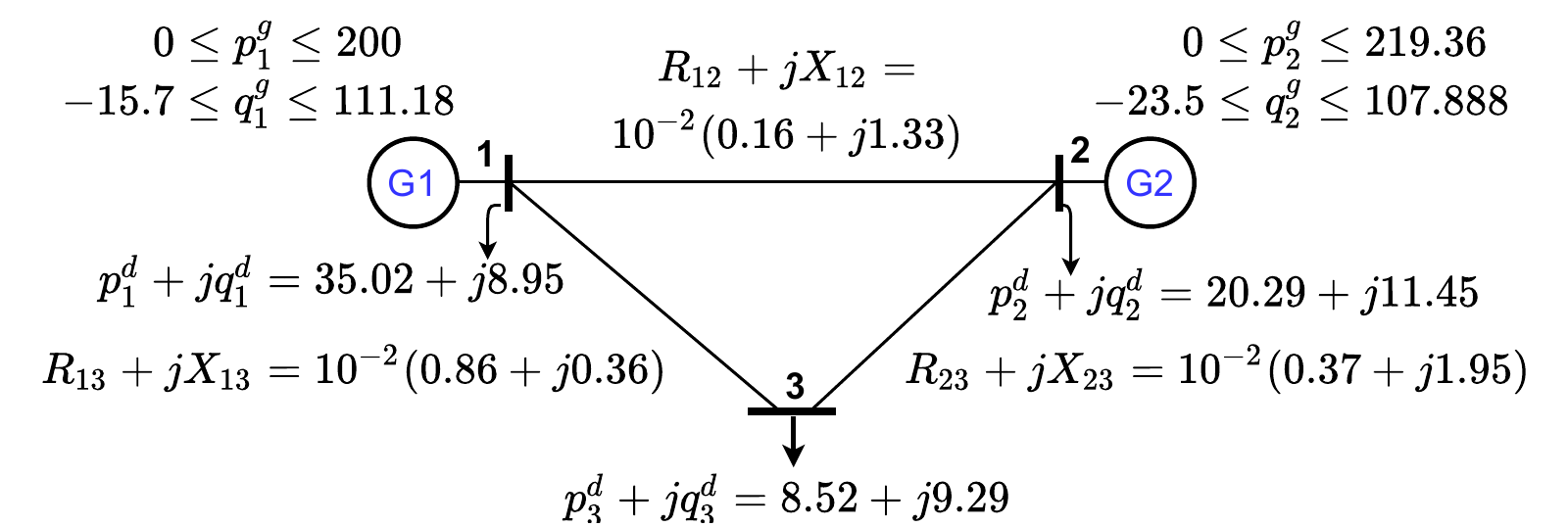}}
	\subfigure[]{\includegraphics[width=0.49\columnwidth]{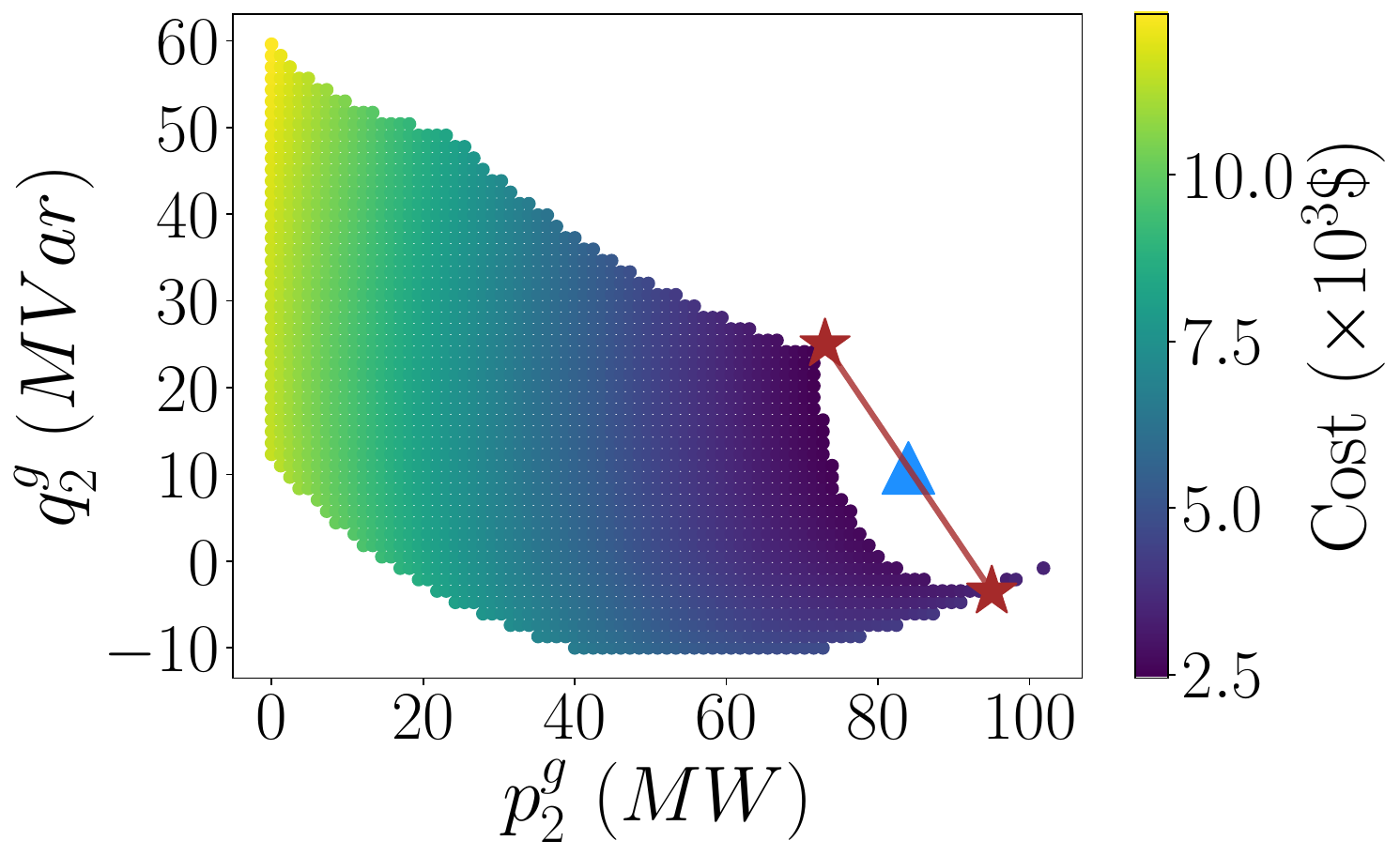}} 
	\subfigure[]{\includegraphics[width=0.49\columnwidth]{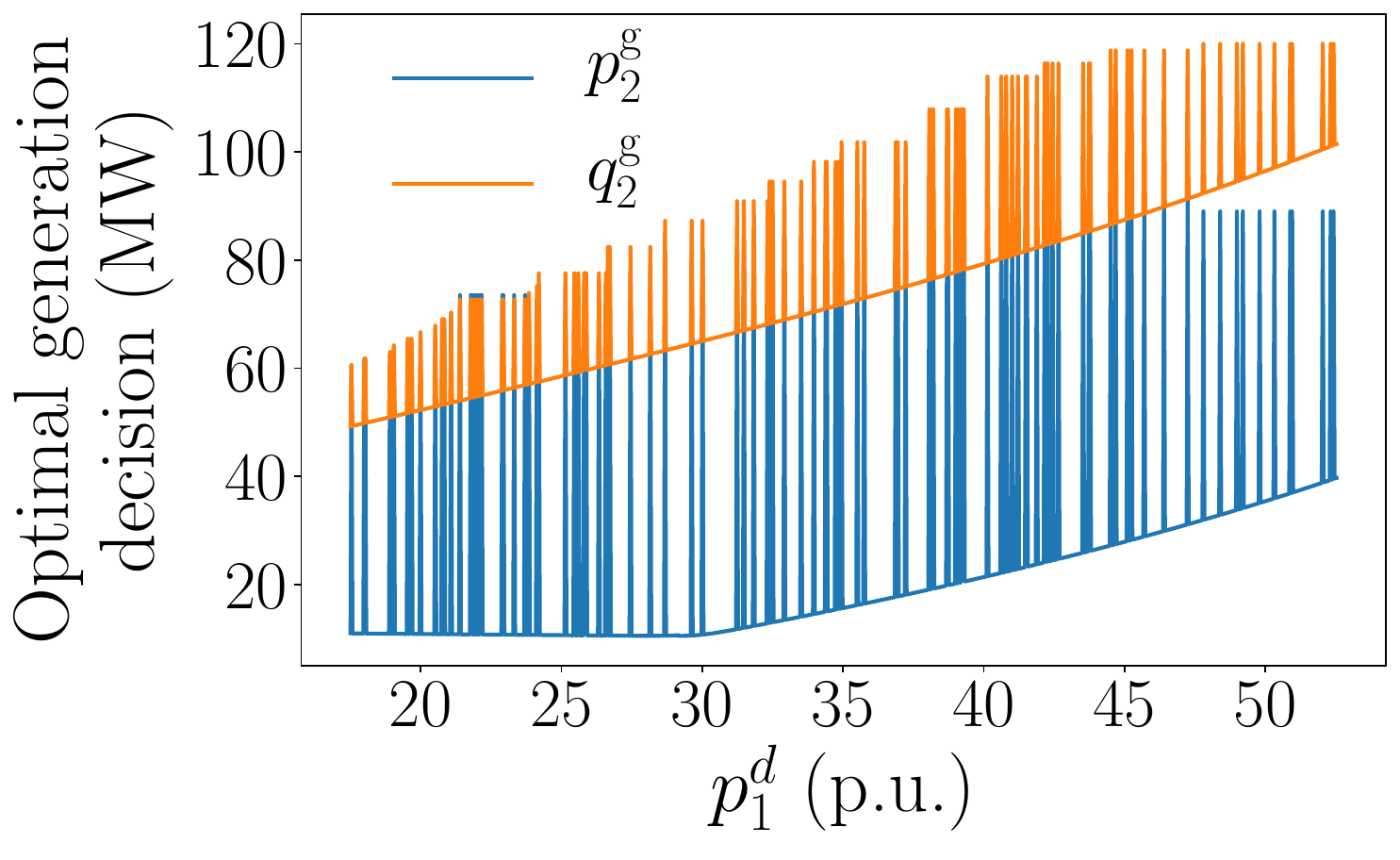}} 
	\subfigure[]{\includegraphics[width=0.49\columnwidth]{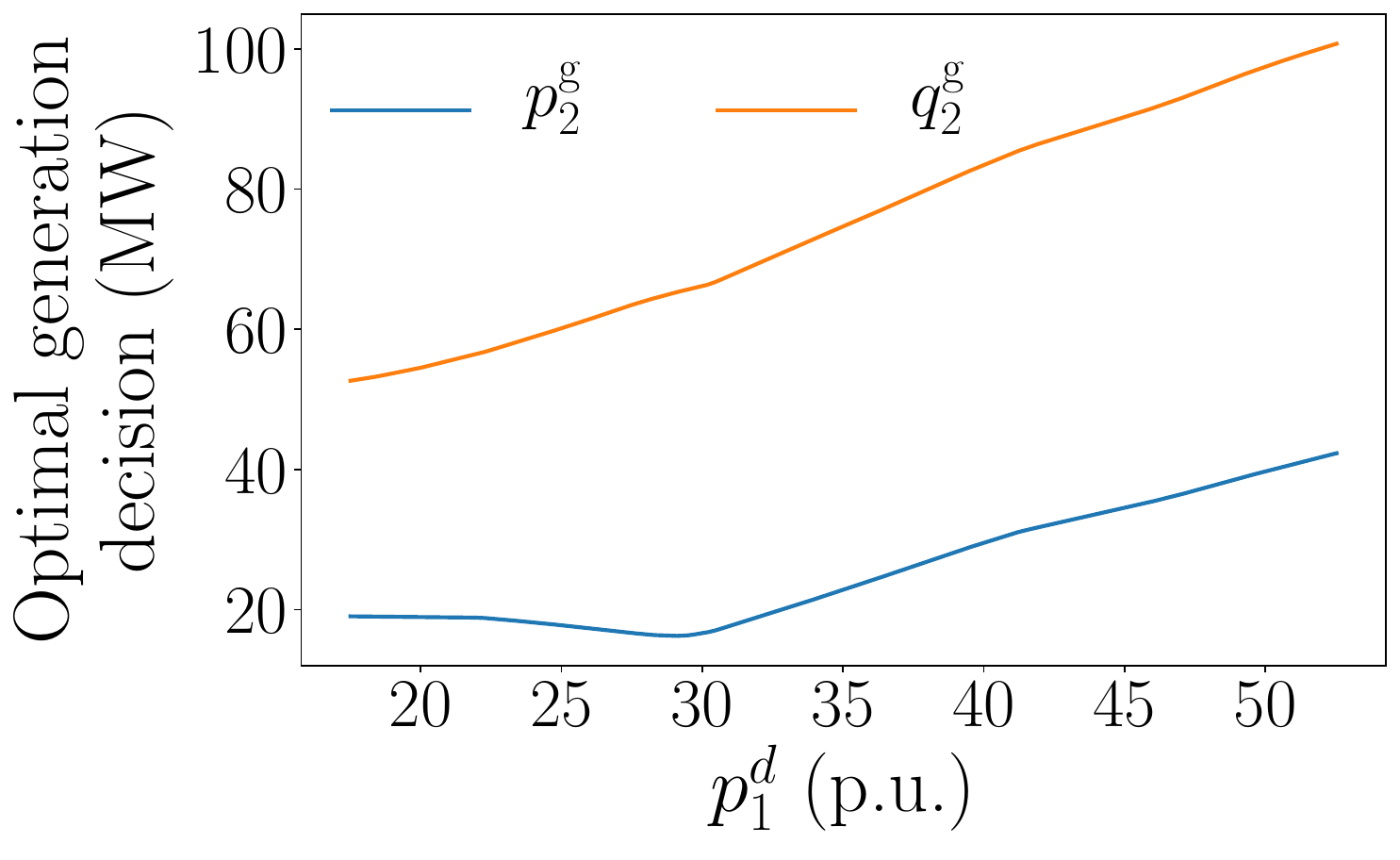}}
	\subfigure[]{\includegraphics[width=0.49\columnwidth]{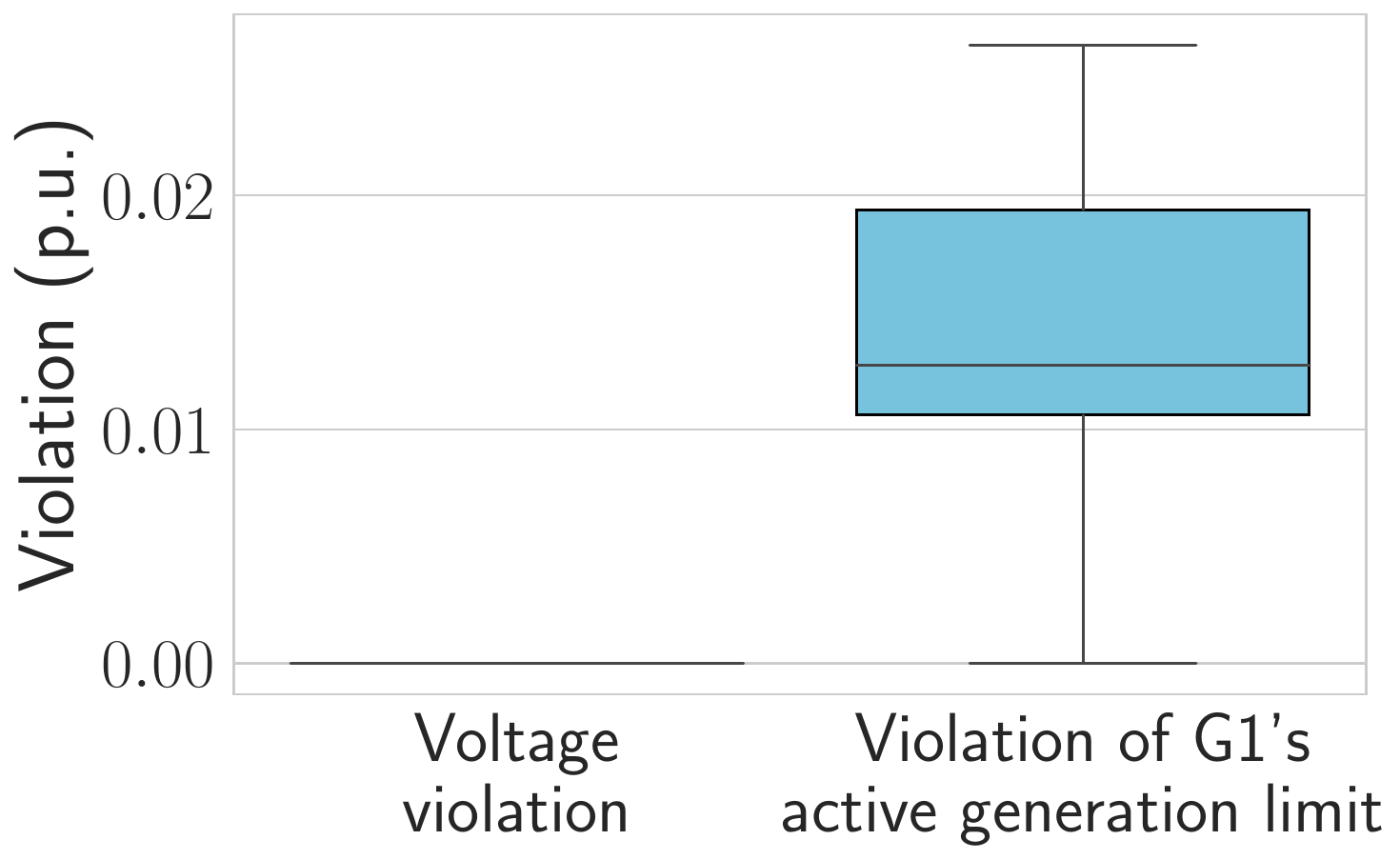}}
	\vspace{-4mm}
 	\caption{An example to illustrate the discontinuity of the mapping from loads to the decisions given by the IPM. (a) Structure of the test system, (b) Feasible set of the second generator's output $(p^g_2, q^g_2)$, where the red stars represents two different local minimums, (c) decisions given by IPM, (d) decisions predicted by a neural network, and (e) constraint violations of the predicted decisions.}
	\label{fig_MSE_2}
	\vspace{-4mm}
\end{figure}

Most existing studies, including \cite{9844847,10023997,9335481}, utilize the solutions of IPM as training labels. These solutions are local minima, while an AC OPF instance may contain multiple local minima due to its non-convex nature. 
Hence, slight changes in loads can shift IPM's solution to a different local minimum, leading to a discontinuous mapping from loads to training labels. Despite this, these studies typically train continuous neural networks with MSE loss as surrogate models. Thus, the predictions of these surrogates may significantly differ from the training labels near points of discontinuity. This discrepancy raises concerns about the feasibility performance of the surrogates. 
To illustrate this feasibility issue, we present an example using a simple 3-bus system. Fig. \ref{fig_MSE_2}(a) displays the system's structure. This example restricts all bus voltages within [0.8 p.u., 1.2 p.u.], while the line flow constraint is ignored. The unit generation costs for the two generators G1 and G2 are 100.5 \$/MWh and 499.8 \$/MWh, respectively.
Fig. \ref{fig_MSE_2}(b) shows the feasible region of the second generator's output $(p^g_2, q^g_2)$, where various colors indicate different total generation costs. This region is non-convex and includes multiple local minima, e.g., the two points marked with red stars. If a small change in the active load $p_1^d$ shifts the IPM solution from one red star to the other, a model trained with MSE loss tends to predict a point between them, like the blue triangle, which is infeasible. 
Fig. \ref{fig_MSE_2}(c) demonstrates the relationship between the active load $p_1^d$ and IPM's solutions, where each curve contains numerous discontinuities. Fig. \ref{fig_MSE_2}(d) shows the predictions of a neural network trained with MSE loss, which visibly differ from the IPM's solutions at discontinuous points. Fig. \ref{fig_MSE_2}(e) illustrates the constraint violations of these predictions. Obviously, these predictions significantly violate the first generator's active power limit. This example indicates that employing MSE loss may lead to a feasibility issue.

\subsection{Decision loss} \label{sec_decisionLoss}
We introduce the ``decision loss" to overcome the previous two challenges. The decision loss use the original objective of the AC-OPF problem as the loss function:
\begin{align}
L^\mathrm{DL}_{\mathbf W}(\mathbf x) = \frac{1}{|D|}\sum_{n \in \mathcal{D}} \text{Cost}\left( \bm \pi(\mathbf x_n; \mathbf W)\right). \label{eqn_decisionLoss}
\end{align}
Compared to the MSE loss, this decision loss offers two specific advantages: {i) It can directly quantify the optimality of decisions, i.e., a decision with a lower decision loss must result in a lower cost than the one with a higher decision loss; ii) The use of decision loss avoids explicitly referring to the training labels from IPM (which may involve the discontinuous mapping mentioned in Section~\ref{sec_MSE}). As to be discussed in Section~\ref{sec_training}, stochastic gradient descent can be used to train AC OPF surrogates with the decision loss together with nonconvex OPF constraints, which may enable us to escape from the local optima\cite{xie2021a}. Therefore, this loss may help mitigate the feasibility issue caused by using MSE. 
%
}


\subsection{Structure of the AC OPF surrogate}
\begin{figure}
		\vspace{-4mm}
	\centering
	{\includegraphics[width=0.9\columnwidth]{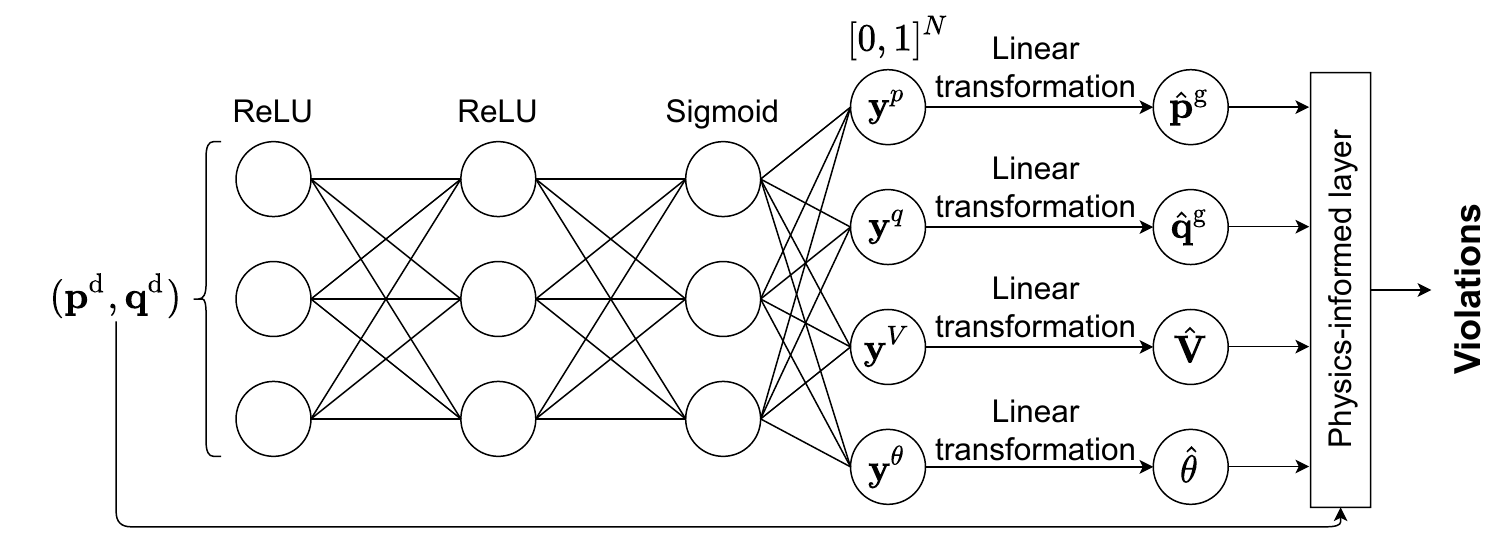}}
	\vspace{-2mm}
 	\caption{Specially designed structure of the neural network for predicting OPF solutions. Its last activation is the sigmoid. A physics-informed layer is also established for the calculation of constraint violations.}
	\label{fig_NN}
	\vspace{-4mm}
\end{figure}

Since the decision loss can not capture OPF constraints, training a surrogate with it directly may lead to poor feasibility. To address this isse, we first design a special structure for this surrogate, as shown in Fig. \ref{fig_NN}. The output of the last hidden layer is composed of four parts: $\mathbf{y}^\mathrm{p} \in \mathbb{R}^N$, $\mathbf{y}^\mathrm{q} \in \mathbb{R}^N$, $\mathbf{y}^\mathrm{V} \in \mathbb{R}^N$, and $\mathbf{y}^\theta \in \mathbb{R}^N$. Since the last activation is set as the sigmoid function, their values are always within $[0,1]^N$. 
A linear transformation is further applied to every part to ensure that the predicted decision can satisfy the voltage and generation limits \eqref{eq:voltage_magnitude}-\eqref{eq:qg_limits}. For instance, the linear transformation for $\mathbf{y}^\mathrm{p} \in \mathbb{R}^N$ is expressed as follows:
\begin{align}
\hat{\mathbf{p}}^\mathrm{g} = \mathbf{y}^\mathrm{p} * (\mathbf{p}^{\mathrm{g,max}}-\mathbf{p}^{\mathrm{g,min}}) + \mathbf{p}^{\mathrm{g,min}},
\end{align}
where $*$ represents element-wise multiplication; $\mathbf{p}^{\mathrm{g,max}}$ and $\mathbf{p}^{\mathrm{g,min}}$ are the vector forms of ${p}_i^{\mathrm{g,min}}$ and ${p}_i^{\mathrm{g,max}}$. Then, the predicted active generation decision $\hat{\mathbf{p}}^\mathrm{g}$ can always satisfy \eqref{eq:pg_limits}. 

A physics-informed layer is further involved for the calculation of constraint violations. Specifically, by substituting the predicted voltage magnitude $\hat{\mathbf{V}}$ and angle $\hat{\bm{\theta}}$ into \eqref{eq:pf_relation} and \eqref{eq:qf_relation}, the corresponding branch power flows, i.e., $\hat{p}_{ij}^\mathrm{f}$ and $\hat{q}_{ij}^\mathrm{f}$, can be obtained. Then, the violation of power flow limit \eqref{eq:line_flow}, i.e., $\sigma_{ij}^\mathrm{f}$, can be calculated by:
\begin{align}
\sigma_{ij}^\mathrm{f} = \max\left\{\sqrt{|\hat{p}_{ij}^\mathrm{f}|^2 + |\hat{q}_{ij}^\mathrm{f}|^2} - s_{ij}^{\mathrm{max}},0\right\}, \ \forall (i,j) \in \mathcal{E}. \label{eqn_vio_f}
\end{align}
By further substituting $\hat{\mathbf{p}}^\mathrm{g}$ and $\hat{\mathbf{q}}^\mathrm{g}$ into \eqref{eq:active_balance} and \eqref{eq:reactive_balance}, we can calculate the violations of nodal power balance constraints:
\begin{align}
& \sigma_{ij}^\mathrm{p} = \left|\hat p^\mathrm{g}_i - p^\mathrm{d}_i - \sum_{j \in N} \hat p^\mathrm{f}_{ij} \right|, \quad \forall (i,j) \in \mathcal{E}, \label{eq:active_balance_vio} \\
& \sigma_{ij}^\mathrm{q} = \left|\hat q^\mathrm{g}_i - q^\mathrm{d}_i - \sum_{j \ in N} \hat q^\mathrm{f}_{ij} \right|, \quad \forall (i,j) \in \mathcal{E}. \label{eq:reactive_balance_vio}
\end{align}
These violations allow us to involve Lagrangian duality to enhance feasibility.

\subsection{Lagrangian duality-incorporated training} \label{sec_training}

We follow \cite{9335481} and introduce the violation-based Lagrangian duality into the training process to further enhance the feasibility of the predicted decision. Specifically, the training of the surrogate can be formulated as the following optimization problem
\begin{align}
\min_{\mathbf W} & \quad \frac{1}{|D|}\sum_{n \in \mathcal{D}} \text{Cost}\left( \bm \pi(\mathbf x_n; \mathbf W)\right), \\
\text{s.t. }& \quad \mathbf g\left(\mathbf{x}_n, \bm \pi(\mathbf x_n; \mathbf W)\right) \leq \mathbf{0}, \quad \quad \quad  \forall n \in \mathcal{D}, \notag \\
& \quad \mathbf h\left(\mathbf{x}_n, \bm \pi(\mathbf x_n; \mathbf W)\right) = \mathbf{0}, \quad \quad \quad \forall n \in \mathcal{D}, \notag
\end{align}
where we use $\mathbf g \leq \mathbf{0}$ and $\mathbf h = \mathbf{0}$ to represent all inequality and equality constraints in the AC OPF problem. Its violation-based Lagrangian relaxation is expressed as:
\begin{align}
\text{LR} (\bm \lambda, \bm \mu) = \min_{\mathbf{W}} & ~\mathcal{L}(\bm \lambda, \bm \mu, \mathbf x, \mathbf{W}) \notag \\
= \min_{\mathbf{W}} &~\frac{1}{|D|}\sum_{n \in \mathcal{D}} \text{Cost}\left( \bm \pi(\mathbf x_n; \mathbf W)\right) \notag \\
& + \sum_{n \in \mathcal{D}} \bm \lambda_n^\intercal \underbrace{\max\left\{\mathbf g\left(\mathbf{x}_n, \bm \pi(\mathbf x_n; \mathbf W)\right), \mathbf 0\right\}}_\text{Violation of ineq.} \notag \\
&+ \sum_{n \in \mathcal{D}} \bm \mu^\intercal_n \underbrace{\mathbf h^\mathrm{abs}\left(\mathbf{x}_n, \bm \pi(\mathbf x_n; \mathbf W)\right)}_\text{Violation of eq.}, \label{eqn_LR}
\end{align}
where $\bm \lambda_n$ and $\bm \mu_n$ are Lagrange multipliers; $\mathbf{h}^{\mathrm{abs}}$ represents the element-wise absolute values of $\mathbf{h}$. When the neural network introduced in Fig. \ref{fig_NN} is employed to predict the dispatch decision, inequality constraints \eqref{eq:voltage_magnitude}-\eqref{eq:qg_limits} can be naturally satisfied. Thus, the violation of inequality constraints only contains $\sigma_{ij}^\mathrm{f}$. The violation of equality constraints includes $\sigma_{ij}^\mathrm{p}$ and $\sigma_{ij}^\mathrm{q}$. For convenience, we separate the multiplier $\bm \mu$ into two parts, i.e., $\bm \mu_n=(\bm \mu_n^\mathrm{p}, \bm \mu_n^\mathrm{q})$.
Its Lagrangian dual is formulated as:
\begin{align}
\text{LD} = \max_{\bm \lambda, \bm \mu} \text{LR} (\bm \lambda, \bm \mu).
\end{align}
Then, we implement the training process by alternately solving the Lagrangian relaxation problem $\text{LR} (\bm \lambda, \bm \mu)$ and its dual $\text{LD}$. The detailed training procedure is summarized in \textbf{Algorithm} \ref{algorithm_train}. In steps 8-10, we update the Lagrange multipliers by solving LR through a subgradient method \cite{nesterov2009primal}.

\begin{algorithm} \label{algorithm_train}
\DontPrintSemicolon
\SetAlgoLined
\SetKwInput{KwInput}{Input}                
\SetKwInput{KwOutput}{Output}              
\SetKwFor{For}{for}{do}{end}            
\SetKwIF{If}{ElseIf}{Else}{if}{then}{else if}{else}{end}

\caption{The training of the OPF surrogate}

\KwInput{Training data $\{\mathbf{x}_n,\forall n \in \mathcal{D}\}$, learning rate $\alpha$, and stepsize $\rho$}

$(\bm \lambda_n, \bm \mu_n) \leftarrow \text{init}, \forall n \in \mathcal{D}$ and $\mathbf{W} \leftarrow \text{init}$ \;
\For{\text{epoch} $k = 1,2,\dots$}{
    $\mathcal{L}(\bm \lambda, \bm \mu, \mathbf x, \mathbf{W}) \leftarrow \text{Eq. \eqref{eqn_LR}}$ \;
    $\sigma_{ij,n}^{\mathrm{f}}  \leftarrow \text{Eq. \eqref{eqn_vio_f}},   \ \forall (i,j) \in \mathcal{E}, \ \forall n \in \mathcal{D}$ \;
    $\sigma_{ij,n}^{\mathrm{p}}  \leftarrow \text{Eq. \eqref{eq:active_balance_vio}},   \ \forall i \in \mathcal{N}, \ \forall n \in \mathcal{D}$ \;
    $\sigma_{ij,n}^{\mathrm{q}}  \leftarrow \text{Eq. \eqref{eq:reactive_balance_vio}},   \ \forall i \in \mathcal{N}, \ \forall n \in \mathcal{D}$ \;
    $\mathbf{W} \leftarrow \mathbf{W} - \alpha \nabla_{\mathbf{W}}\mathcal{L}(\bm \lambda, \bm \mu, \mathbf x, \mathbf{W})$ \;
    $\lambda_{ij, n} \leftarrow \lambda_{ij, n} + \rho \cdot \sigma_{ij,n}^{\mathrm{f}},   \ \forall (i,j) \in \mathcal{E}, \ \forall n \in \mathcal{D} $ \;
    $\mu_{ij, n}^{\mathrm{p}} \leftarrow \mu_{ij, n}^{\mathrm{p}} + \rho \cdot \sigma_{ij,n}^{\mathrm{p}},   \ \forall i \in \mathcal{N}, \ \forall n \in \mathcal{D} $ \;
    $\mu_{ij, n}^{\mathrm{q}} \leftarrow \mu_{ij, n}^{\mathrm{q}} + \rho \cdot \sigma_{ij,n}^{\mathrm{q}},   \ \forall i \in \mathcal{N}, \ \forall n \in \mathcal{D} $ \;
}
\KwOutput{Trained model $\bm \pi(\mathbf x; \mathbf W)$}
\end{algorithm}

\section{Case study} \label{sec_case}

\subsection{Simulation setting up}
We implement a case study based on the IEEE 39-bus test system to verify the benefits of the proposed decision loss. 
This test system contains 10 generators and 46 lines. Its voltage level is 345KV, and all bus voltages are restricted within [0.93 p.u., 1.07 p.u.]. Other parameters can be founded in \cite{4113518}.

Two models are implemented for comparison:
\begin{itemize}
\item $\bm \pi^\mathrm{Decision}$: The OPF surrogate trained by the decision loss.
\item $\bm \pi^\mathrm{MSE}$: The OPF surrogate trained by the MSE.
\end{itemize}
Both models are implemented by a neural network with three hidden layers and 60 neurons in each. {Moreover, the Lagrangian duality-incorporated training method introduced in Section \ref{sec_training} is applied to not only $\bm \pi^\mathrm{Decision}$ but also $\bm \pi^\mathrm{MSE}$ to ensure a fair comparison of the results.} 
Since they require historical data for training, we employ Pandapower, a power system simulation toolbox in Python \cite{8344496}, to construct  a training set. First, 1,000 samples of load conditions, i.e., $\{\mathbf{x}_n, \forall n \in \mathcal{D}\}$, are generated with a uniform distribution as training features. These samples are given to Pandapower, and Pandapower solves AC OPF with the IPM for each sample. Then, the optimal dispatch decision $\{(\mathbf{p}^{g,*}_n,\mathbf{q}^{g,*}_n,\mathbf{V}^{*}_n, \bm{\theta}^{*}_n),  \forall n \in \mathcal{D}\}$ can be obtained as training labels. During training, 80\% of samples are used as training set, while the rest 20\% are regarded as testing set to test the performance of different models.

All numerical experiments are conducted on an Intel(R) 8700 3.20GHz CPU equipped with 16 GB of memory. Both models are implemented and trained using Pytorch.

\subsection{Optimality and feasibility}
\begin{figure}
		\vspace{-4mm}
	\subfigbottomskip=-4pt
	\subfigcapskip=-4pt
	\centering
	\subfigure[]{\includegraphics[width=0.49\columnwidth]{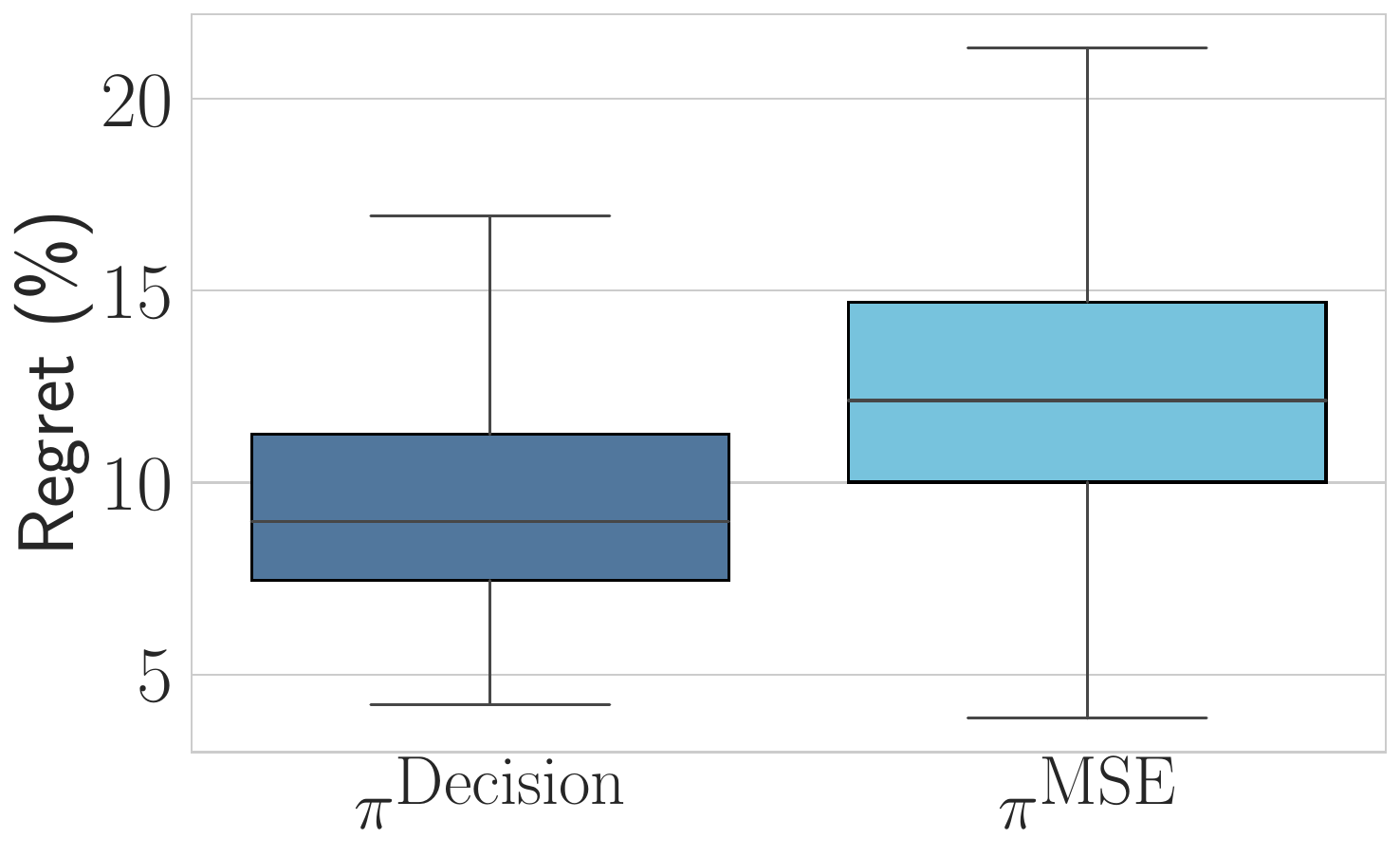}} 
	\subfigure[]{\includegraphics[width=0.49\columnwidth]{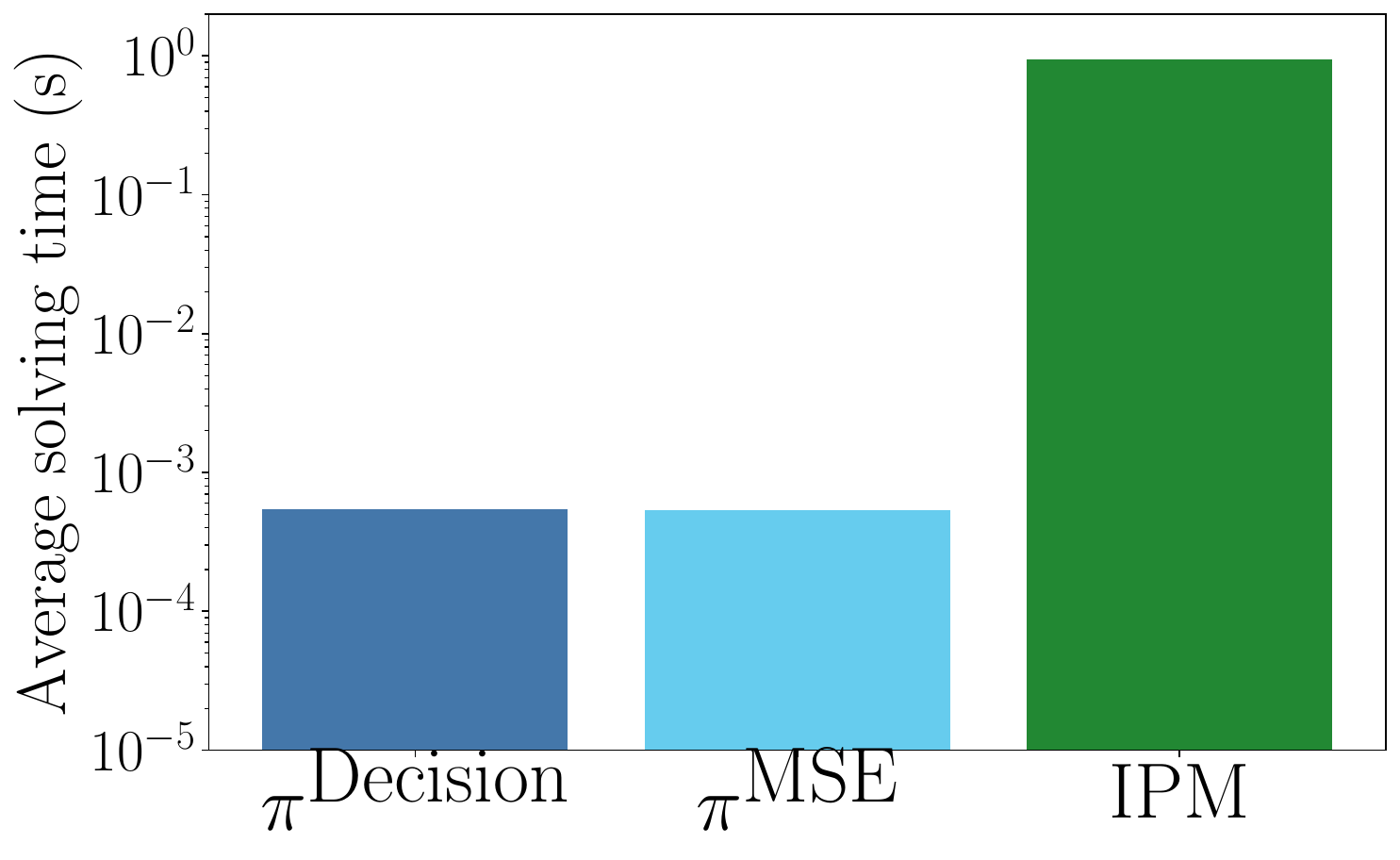}}
 	\caption{Results of (a) regrets and (b) solving time given by the two models. The regret can be regarded as the ``optimality gap". }
	\label{fig_regret}
	\vspace{-4mm}
\end{figure}

Fig. \ref{fig_regret} presents the regret and average solving times for the two test models. Regret is defined as the optimality gap between the surrogate's decision and that given by IPM. Its value may be negative if the surrogate's decisions violate constraints. The surrogate model trained by the decision loss, denoted as $\bm \pi^\mathrm{Decision}$, always shows lower regrets than the one trained by the MSE, i.e., $\bm \pi^\mathrm{MSE}$. As introduced in Section \ref{sec_decisionLoss}, {decision loss can accurately measure the optimality of decisions that the MSE, so $\bm \pi^\mathrm{Decision}$ achieves a lower cost compared to $\bm \pi^\mathrm{MSE}$.} {Since both models replace the solving procedure with the forward pass of neural networks, they can output dispatch decisions instantly. Hence, the solving times, i.e., the time needed to output the optimal dispatch for a new load condition, are three orders of magnitudes smaller than that of IPM.} These results confirm that the decision loss can achieve better optimality compared to the MSE.

\begin{figure}
	\centering
	{\includegraphics[width=0.9\columnwidth]{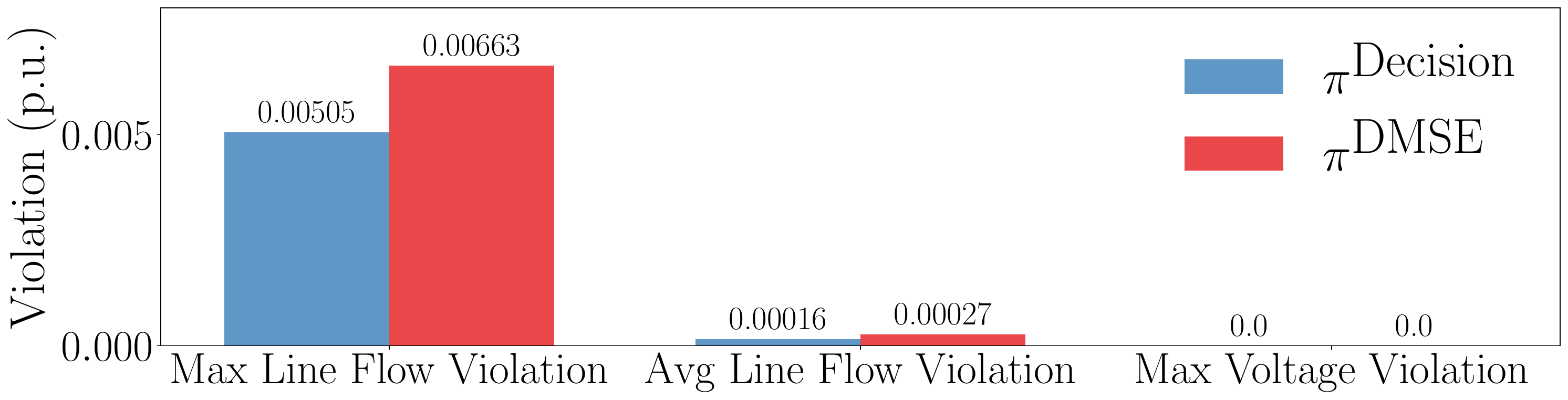}}
	\vspace{-4mm}
 	\caption{Maximum and average line power flows and maximum bus voltage violations of different methods.}
	\label{fig_vio_39Bus}
	\vspace{-4mm}
\end{figure}

Fig. \ref{fig_vio_39Bus} displays the constraint violation results for the two models. The line flow violations of both models are very small. Moreover, the bus voltage limit can always be satisfied. These results validate the effectiveness of the special neural network structure and the Lagrangian duality-incorporated training method in improving feasibility. The violations of $\bm \pi^\mathrm{Decision}$ are slightly smaller than those of $\bm \pi^\mathrm{MSE}$, which indicates that using the decision loss can mitigate the feasibility issue caused by the discontinuity in the target mapping.  

\subsection{Effects of neuron numbers}

\begin{figure}
		\vspace{-4mm}
	\subfigbottomskip=-4pt
	\subfigcapskip=-4pt
	\centering
	\subfigure[]{\includegraphics[width=0.49\columnwidth]{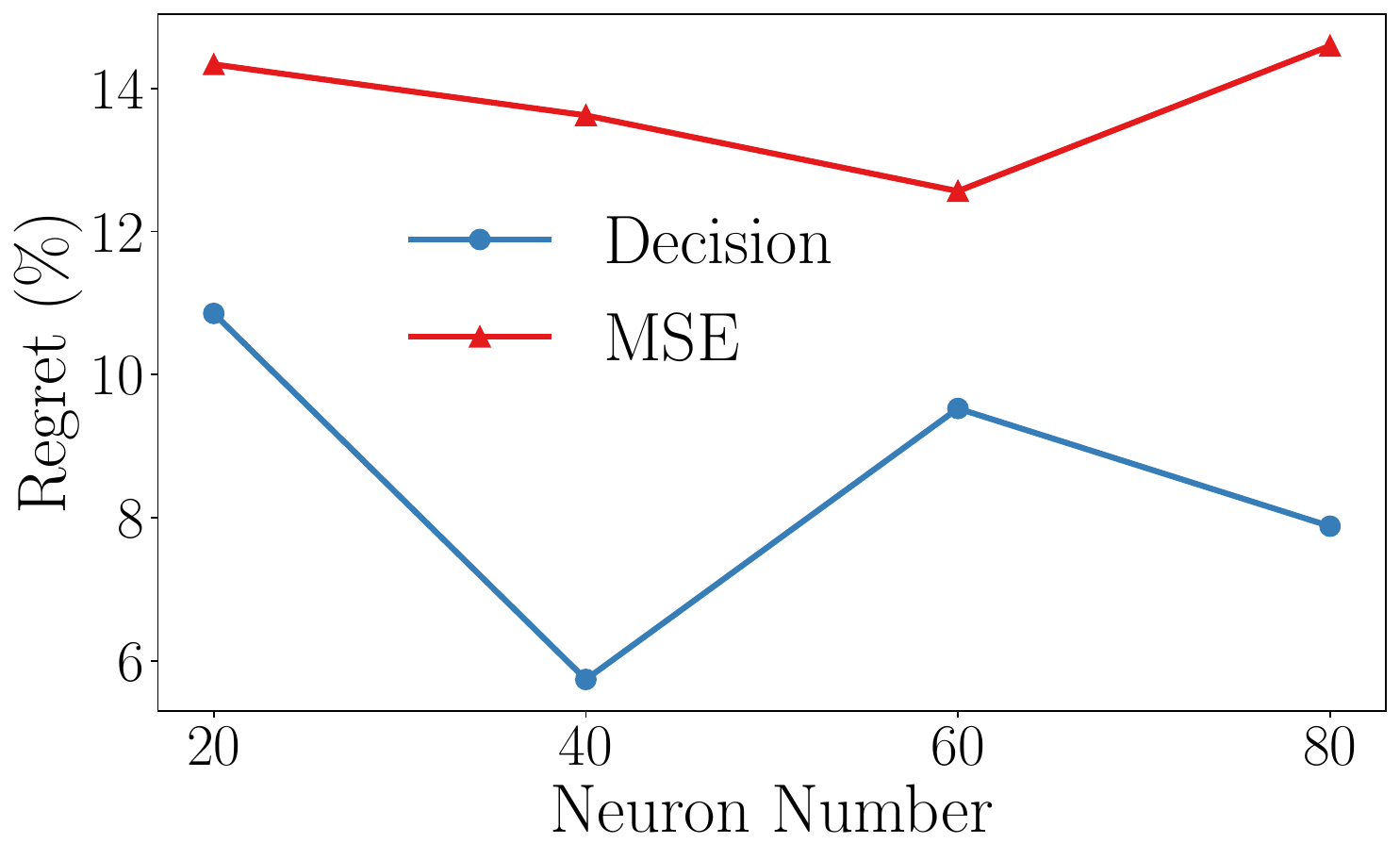}}
	\subfigure[]{\includegraphics[width=0.49\columnwidth]{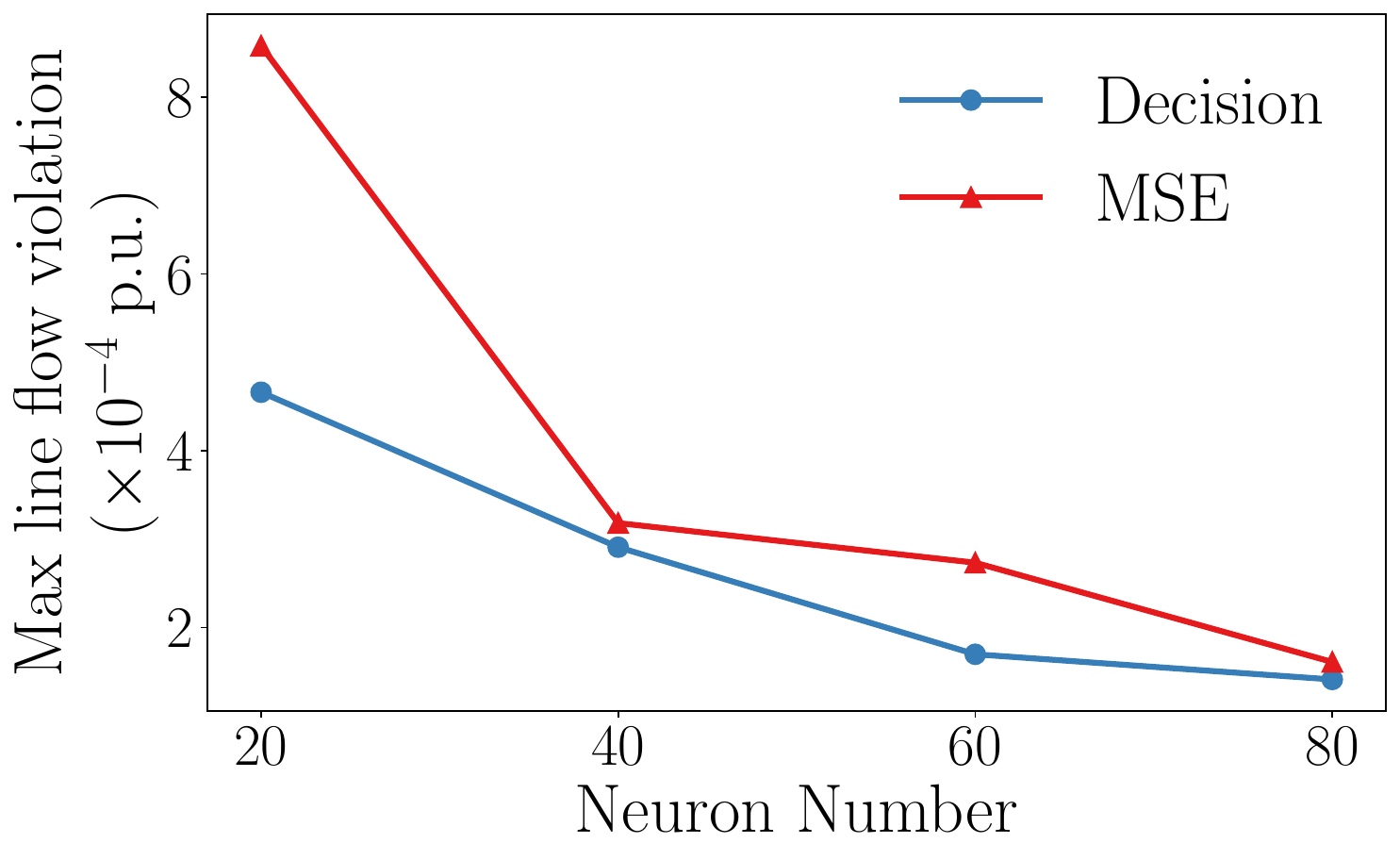}}
 	\caption{Results of (a) average regrets and (b) average line flow violations.}
	\label{fig_results_neuronNum}
	\vspace{-4mm}
\end{figure}

Fig. \ref{fig_results_neuronNum} shows the effects of increasing the neuron number in each hidden layer on the performance of two models. While the average regrets for both models do not consistently decrease with more neurons, the maximum line flow violations decrease substantially. A neural network with more neurons has a stronger capacity for representation, so its training loss can be smaller compared to the ones with fewer neurons. Note the training loss equals the summation of the decision/MSE loss and the penalty for constraint violations. During extensive training over thousands of epochs, the Lagrangian multipliers can increase significantly, so the violation penalty may become the dominant component of the training loss. As a result, increasing the neuron number may not consistently lower the decision loss/MSE, but it does markedly reduce the violation penalty. In each case,  model $\bm \pi^\mathrm{Decision}$ demonstrates lower regret and maximum line flow violation than $\bm \pi^\mathrm{MSE}$, which confirms the superiority of the decision loss over the commonly used MSE in achieving better optimality and feasibility.

\section{Conclusions} \label{sec_conclusion}

This paper compares MSE and decision loss as loss functions in training learning-based OPF surrogates. {We first outline the common procedure of training surrogates using the MSE, and discuss the corresponding potential optimality and feasibility issues. We then introduce the decision loss, which aligns with the OPF objective but does not capture the OPF constraints. Although it can mitigate the issues caused by the MSE, it is not guaranteed to output feasible solutions.} To overcome this, we further develop a specially structured neural network and incorporate Lagrangian duality into the training process to improve feasibility. Simulations on the IEEE 39-bus test system demonstrate that decision loss outperforms MSE in both optimality and feasibility.



\footnotesize
\bibliographystyle{ieeetr}
\bibliography{ref}
\end{document}